# Immersive Augmented Reality Training for Complex Manufacturing Scenarios

Mar Gonzalez-Franco[1]*, Julio Cermeron[2], Katie Li[1,2], Rodrigo Pizarro[1,3], Jacob Thorn[1], Paul Hannah[1], Windo Hutabarat[2], Ashutosh Tiwari[2], Pablo Bermell-Garcia[1]

[1] Airbus Group Innovations UK (pablo.bermell@airbus.com)
[2] Manufacturing Department, Cranfield University UK
[3] Experimental Virtual Environments for Neuroscience and Technologies Lab (EVENTLab), Universitat de Barcelona

**ABSTRACT**

In the complex manufacturing sector a considerable amount of resources are focused on developing new skills and training workers. In that context, increasing the effectiveness of those processes and reducing the investment required is an outstanding issue. In this paper we present an experiment that shows how modern Human Computer Interaction (HCI) metaphors such as collaborative mixed-reality can be used to transmit procedural knowledge and could eventually replace other forms of face-to-face training. We implement a real-time Immersive Augmented Reality (IAR) setup with see-through cameras that allows for collaborative interactions that can simulate conventional forms of training. The obtained results indicate that people who took the IAR training achieved the same performance than people in the conventional face-to-face training condition. These results, their implications for future training and the use of HCI paradigms in this context are discussed in this paper.

**INTRODUCTION**

Nowadays industry and mass manufacturing in particular are either robotized or rely heavily on skilled workers. Modern assembly lines are based on high value manufacturing and, in that context, training new workers in complex tasks is an outstanding challenge for the industry [1], as it involves having to dedicate limited physical equipment and professionals to instruct new personnel [2]. Furthermore the operation of dangerous equipment can give rise to health and safety concerns [3]. In this context the use of novel technologies and HCI metaphors to train future workers on the processes could both increase the safety and reduce the training costs of future workers, which translate to an increase in productivity.

Up to now, several computer-based approaches have been used as alternative methods for reducing the impact of these hurdles in industrial training. Previous work includes the use of Virtual Environments which allow users to practice and rehearse situations that might otherwise be dangerous in a real environment [4]. These approaches have been used for training in a variety of disciplines including health and safety [5], [6], medical training [7], [8], and fire services [4], among others. Research has shown how learning and performance can be improved with these technologies [9]. However, most computer-based training systems do not reproduce with enough fidelity scenarios existing in complex assembly training so that they could on the future totally replace conventional physically-based trainings. A prominent characteristic of assembly training is its collaborative component where face-to-face interaction seems to play a great role for learning [10]. Furthermore, in real deployment workers have access to physical equipment which they manipulate on demand, there is a need to make the digital training tangible and achieve similar levels of fidelity. Indeed, our goal is to reproduce that experience as much as possible. To achieve those levels of natural interaction we turn into Virtual Reality (VR), where it has been shown that objects can be manipulated naturally and from a first person perspective when the participants position and movements are tracked [11], [12].

VR applications are especially powerful when participants experience the presence illusion: the feeling of actually "being there" inside the simulation. In fact, presence has been described by a combination of two factors: the plausibility of the events happening being real, and the place illusion, the sensation of being transported to a new location [13], [14]. These illusions, especially when combined, can produce realistic behaviors from participants [15]. Indeed VR has been used by social psychologist to find out how people would react in extreme situations, such as violent scenarios [16] or even to reproduce moral dilemmas to find out how people react without compromising their integrity [17].

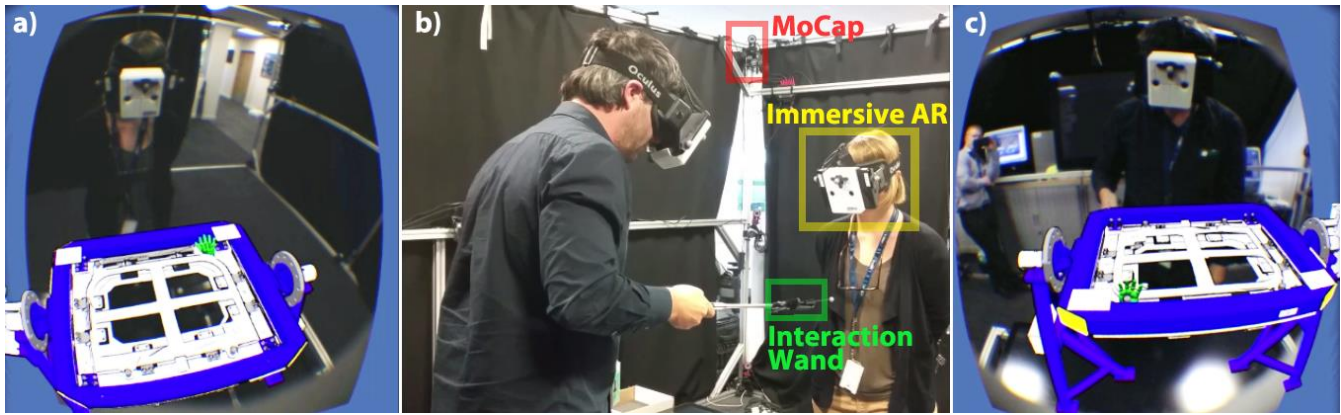

**Figure 1. Immersive AR setup. (A) Trainer's view, see-through with the virtual assembly jig. (B) Laboratory equipped with 24 motion capture cameras and two participants wearing the Immersive AR setups set for collaboration: the trainer is carrying the interaction wand while the second person observes the operation (C). The Interaction wand in is represented by the green actuator.**

Of particular interest when providing VR with Head Mounted Displays (HMDs) is the self-body experience, since a strong body ownership can be generated with the proper multisensory integration [18]. When a virtual body is experienced from a first person perspective [19], [20] and participants have control over that body, agency, this creates the illusion that the real body has been substituted by the virtual one [21]–[23]. In fact, this illusion is so deep into our brain that it produces neurophysiological signatures [20], [23]. Several research on the use of body ownership illusions have shown that they can produce changes in participants' behavior during the VR experiences, for example changing implicit racial attitudes after being immersed [24] or even inducing the feeling of traveling in time [25]. These findings suggest that training scenarios could benefit from using Immersive VR.

Indeed, several authors have already used VR as a tool for training and rehearsal in medical situations [26], [27] and disaster relief training [28] among other fields. However, while VR may be an excellent approach for isolated training, they are increasingly complex to use for collaborative training [29]–[32]: systems require several computers, complex network synchronization and labor-intensive application development. Furthermore, aspects of self-representation and virtual body tracking become of major importance, as to collaborate and communicate in face-to-face scenarios we usually turn to body language.

One approach to overcome the self-representation issue, is to use a sort of augmented reality or mixed-reality paradigms where participants are not only able to explore a digital object from a first person perspective but also to see themselves collocated with real objects and people [33]. This paradigm can be much better suited for many collaborative applications and more in particular for training scenarios where instructor and trainee are together in the same space, and not remotely located. With this technology, a high degree of presence, body ownership and agency is ensured because participants are seeing at least both the real world and their own bodies moving as they move. Additionally, in the training scenario participants can see the instructor guiding them through the process, but without the possible physical harm of the real operation. Finally, network synchronization is greatly simplified, since there is no need to share participants' body tracking information and avatars.

In this paper we validate whether such a mixed reality setup could work in real training scenarios. We present a method to train future operators of expensive and heavy equipment with an IAR application, and we compare the results to a training done face-to-face on a physical scaled model of that same hardware.

## MATERIALS AND METHODS

### Participants

Twenty-four volunteers (age mean=32.5, SD=9.6 years old, 3 female) participated in the user study. Due to the confidential nature of the manufacturing content, this study was conducted using only employees from the institution. Participants of the study did not have previous manufacturing knowledge and were asked to complete a demographic questionnaire before participating. This study was approved by the Science and Engineering Research Ethics Committee (SEREC) of Cranfield University. Following the Declaration of Helsinki all participants were given an information sheet, signed informed consent and agreed to participate in the study.

### Procedure

The experiment implemented two different methods of training: (i) conventional training, where participants were taught in a face-to-face scenario manipulating a real assembly jig, and (ii) Immersive AR training, where participants were taught in a face-to-face scenario with a see-through augmented reality setup that allowed collaboration over a rendered digital model of the assembly jig, this setup also implemented the manipulations and interactions in real-time necessary for the training. In both conditions participants underwent the same procedural script obtained from a complex manufacturing manual of an aircraft maintenance door. Participants were then evaluated to assess how much knowledge they captured during the training (Figure 2).

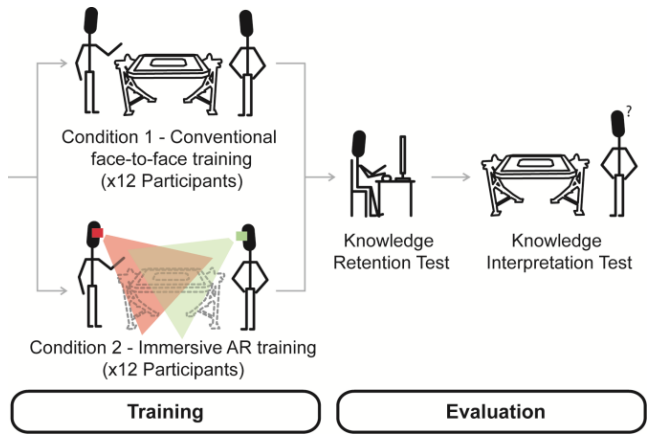

Figure 2. Experimental design and procedure.

Participants were randomly assigned to one of the two experimental conditions in a between subjects study and underwent the following phases after completing the demographic questionnaire:

1. *Training*. The trainer followed a procedural manual to perform the inspection and to operate the moving parts of a door assembly jig. During this phase the trainee has to observe what the trainer is doing and tries to remember as much as possible for the evaluation phase.

2. *Evaluation*. Just after the training, the trainee was asked to complete two test (a knowledge retention and knowledge interpretation test) to compare both types of trainings. The *knowledge retention test* was a written test using a multiple choice format with 8 questions (Table 1). This test was designed to evaluate how much factual knowledge was retained from the training [6]. The *knowledge interpretation test* evaluated whether the procedure of the assembly was properly captured. This test was executed in a scaled physical jig and the trainee was asked to perform step by step significant parts of the assembly training (See supplementary video).

### Knowledge Retention Questions

1. How would you know what PPE you will need?
2. What PPE do you need to wear?
3. What do you need to do during the X operation?
4. To prepare the jig to X, what do you need to X?
5. How many X are needed to secure the X?
6. How many X are needed to secure the X?
7. What do you have to do before X?
8. How do you fit the X?

Table 1. Questions of the Knowledge Retention test. The test had a multiple choice format.

**Apparatus**

In the Immersive Augmented Reality (IAR) condition, participants were donned an Oculus Rift DK1 HMD with a 1280x800 resolution (640x800 per eye), a 110º diagonal Field Of View (FOV) and approximately 90º horizontal FOV. The HMD mounted a pair of cameras to form a see-through IAR setup as in [33], [34]. The scenario was implemented in Unity3D, and the head tracking was performed with a NaturalPoint Motive motion capture system (24 x Flex 13 cameras) running at 120Hz and streaming the head's position and rotation to our application. With this information, we could display the virtual objects from a first person perspective [12]. To interact with the virtual jig, we attached a rigid body marker to an ipow z07-5 stick that contains a button and bluetooth communication; this way participants can view a virtual object matching the position of the marker and press the button to perform actions in the virtual jig (see supplementary video). This Immersive AR system allowed multi-user collaborations where different participants had their head tracked and could interact with each other's virtual objects through a PhotonServer installed in the laboratory (Figure 1, see supplementary video). For the conventional training condition and the participants' Knowledge Interpretation evaluation of both conditions, we manufactured a laser-cut physical model of the jig in transparent plastic (see supplementary video).

## RESULTS

### Knowledge Capture

No significant differences were found for knowledge retention (scores from 0 to 8) between the two conditions (Kruskal-Wallis rank sum test $\chi^2(1)= 0.1$, $p=0.7$). The score for the Virtual condition was (M=3.75, SD=1.21), and the score for the Physical condition was (M=3.91, SD=1.44). Both methods of training were providing similar level of factual knowledge, even if this was not very high, given that the maximal score was 8, and participants in both methods were below that score (Figure 3).

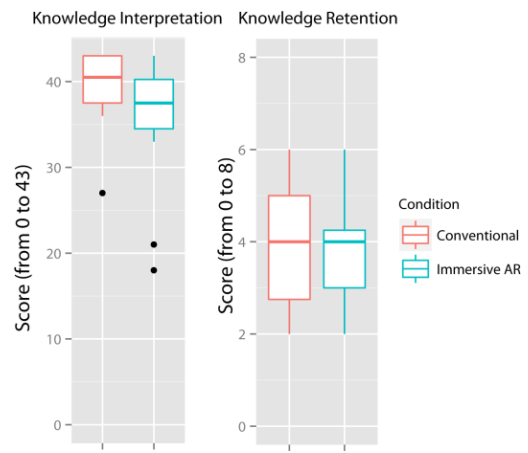

Figure 3. Knowledge Retention and Interpretation scores for both Conventional and IAR conditions.

In the case of knowledge interpretation we find both methods to be performing similarly well. No significant differences were found for knowledge retention (scores from 0 to 43) between the two conditions (Kruskal-Wallis rank sum test $\chi^2(1)= 1.9$, p=0.16). The score for the Virtual condition was (M=35.41, SD=8.03), and the score for the Physical condition was (M=39.25, SD=4.86). Given the high score for both conditions, the procedural training can be considered successful (Figure 3).

When studying the relation of both kinds of knowledge capture we find that while in the Immersive AR condition a correlation trend was found between high scores in the Interpretation and Retention (Pearson $r(12)=0.57$, p=0.052), this was not true for the Conventional training condition (p>0.39) (Figure 4). Moreover, it seems that participants that were performing well in the Immersive AR condition were as good as the ones in the conventional training. It could be that the low performer participants in the Immersive AR were overwhelmed by the setup and that constrained their capacity to capture knowledge, we hypothesize that this effect might fade away as participants become more used to the technology itself.

**Time**

The time spent to complete the training was significantly higher in the Immersive AR condition (M=12.1, SD=2.5 minutes) than in the Conventional training condition (M=9.9, SD=0.9 minutes) (Kruskal-Wallis rank sum test $\chi^2(1)= 0.64$, p=0.01) (Figure 4). This could be partially due to the extra hassle of fitting the equipment, which required time to familiarize with the interaction metaphors and the novelty of the Immersive AR setup.

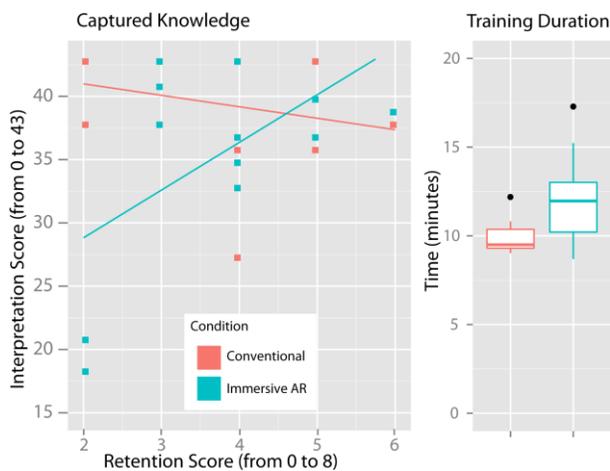

**Figure 4. Left) Correlation between Knowledge Retention and Interpretation scores for both IAR and Conventional conditions. Rigth) training duration for both conditions.**

**DISCUSSION**

Overall, we find that both in the mixed-reality setup and in the face-to-face setup participants acquired similar levels of knowledge. Very high scores were found in the interpretation test, scoring over 80% of accuracy with a single training session in a task that was totally novel to all of them. These results validate our training methodology which was a practical example of a complex manufacturing procedure. More importantly, participants of the mixed-reality scenario achieving results equivalent to those of the physical training, shows that the mixed-reality scenario has provided a successful metaphor for collaborative training. We did however find a correlation between high interpretation and retention scores in participants performed who completed the training through mixed-reality, such correlation was not found with the physical training results. These results are aligned with previous studies that show higher cognitive load is needed when using novel technologies at first [35], and it could show that people in the mixed reality might need to retain more information to move forward in the training. This could be related to the fact that in the mixed-reality setup participants are placed outside their comfort zone, making them unable to remember or guess what to do next. This would also contribute towards explaining the results that show that participants took longer in the VR training than in the physical, because they were less familiar with the environment. Nevertheless, the actual post-training knowledge scores were not significantly different between participants the mixed-reality training and the physical one, thus evidencing the great possibilities in the use of mixed-reality for complex manufacturing training. We hypothesize that these positive results are closely linked to the theories of self-representation and first person interaction with the digital object, which are borrowed from previous studies in VR [36].

**CONCLUSION**

The current paper has presented and validated the use of mixed-reality metaphors for complex manufacturing training by running a user study and measuring the post-training knowledge retention and interpretation scores. The results show equivalent knowledge retention and interpretation for the mixed-reality training and the conventional face-to-face physically based training. These results support the idea that mixed-reality setups can achieve high performances in the context of collaborative training. The implications of these results are clear for the manufacturing industry, but also for the HCI community as it shows evidence of how the integration of existing metaphors for collaborative work and interaction from VR can be implemented in Immersive Augmented Reality.

**ACKNOWLEDGMENTS**

This study was funded with an Innovate UK grant (Technology Strategy Board project:16841-120195).